\documentclass[prd,superscriptaddress,nofootinbib,amsmath,amssymb,aps,11pt]{revtex4-2}
\usepackage[utf8]{inputenc}

\usepackage{bm}
\usepackage{amsfonts}
\usepackage{latexsym}
\usepackage{graphicx}
\usepackage{amsmath}
\usepackage{anyfontsize}
\usepackage{palatino}
\usepackage{mathpazo}
\usepackage[normalem]{ulem}
\usepackage{soul}
\usepackage{comment}

\usepackage{xcolor}     

\begin{document}
\title{\boldmath The spooky ghost of vectorization}

\author{Lorenzo Pizzuti}
\email{lorenzo.pizzuti@unimib.it}
\affiliation{Dipartimento di Fisica G. Occhialini, Universit\'a degli Studi di Milano Bicocca, Piazza della Scienza 3, I-20126 Milano, Italy}

\author{Alexandre M. Pombo}
\email{pombo@fzu.cz}
\affiliation{CEICO, Institute of Physics of the Czech Academy of Sciences, Na Slovance 2, 182 21 Praha 8, Czechia}

\begin{abstract}
    An interesting mechanism for the formation of hairy black holes occurs when a vector field, non-minimally coupled to a source term, grows from a perturbation of the vacuum black hole, \textit{aka} vectorization. Its study has, however, been lacking, in part due to the constant threat of ghost instabilities that have plagued vector fields. In this work, we show evidence that, in a generic family of extended-vector-tensor theories where the vector field is non-minimally coupled to the model's invariant (source term), a spherically symmetric, vectorized black hole always suffers from ghost instabilities. These ultimately turn the process of vectorization astrophysically unviable.
\end{abstract}

\maketitle

%
\section{Introduction}\label{S1}
%
    Never in the more than 100 years of general relativity (GR) has been a better time to study compact objects. The gravitational wave emission from the binary coalescence of black holes (BHs) detected by the LIGO-VIRGO collaboration (\textit{e.g.}~\cite{abbott2016observation,abbott2021gwtc}) and the direct imaging by the Event Horizon Telescope~\cite{collaboration2019first,akiyama2022first}, led to one of the most significant advances in BH history, allowing the study of gravity in its strong field regime where deviations from GR may arise. 

    One of the simplest and more attractive alternatives to GR, which has been extensively studied both at the astrophysical (\textit{e.g.} \cite{Tuna22,khodadi2020black,berti2015testing}) and cosmological (\textit{e.g.} \cite{Kobayashi:2018xvr,Burrage19,Pizzuti22,Ballardini23}) level, consists on the addition of a scalar field to GR. 

	Of particular interest are theories where the scalar field non-minimally couples to an invariant of the theory. These are known as extended-scalar-tensor theories (eST) \cite{crisostomi2016extended}, where BH's scalarization may arise (see \cite{doneva2022scalarization} for a review). 

    Scalarization occurs when perturbations of the vacuum solution push the BH to transfer part of its energy density to a surrounding scalar hair, giving rise to new BH solutions with significant deviations from the vacuum GR counterpart.

	Two exemplary models for the source term able to scalarize are the Gauss-Bonnet~\cite{antoniou2018evasion,doneva2018new,silva2018spontaneous} and Maxwell~\cite{herdeiro2018spontaneous} invariant. While the former, is more astrophysically interesting \cite{doneva2022dynamical}, the latter is relatively simpler and easier to compute\footnote{In a dynamical astrophysical environment, the presence of plasmas around the BH leads to prompt discharge. Alternatively, the neutralization can occur through Hawking charge evaporation \cite{gibbons1975vacuum}.}, serving as a toy model for a wide variety of coupling functions and dynamical studies. In both cases, perturbatively and entropically stable solutions can be formed dynamically by either a linear (\textit{aka} spontaneous/normal scalarization~\cite{fernandes2019spontaneous,astefanesei2019einstein,fernandes2019charged,herdeiro2021aspects,brihaye2019spontaneous,jiang2020spontaneous,myung2019instability,myung2019quasinormal}) or a non-linear~\cite{BlazquezSalcedo:2020nhs,LuisBlazquezSalcedo:2020rqp,doneva2022beyond,pombo2023effects} perturbation of the vacuum solution\footnote{Similar studies were performed for magnetized BHs~\cite{annulli2022spin}, spinning BHs~\cite{hod2020onset,dima2020spin,herdeiro2021spin,berti2021spin,collodel2020spinning} and spinning and charged BH~\cite{herdeiro2021aspects,annulli2022spin}.}. 

    It seems then reasonable to further generalize scalarization to higher particle spins, with a vector field as the most natural first candidate. In fact, spontaneous vectorization~\cite{ramazanouglu2017spontaneous,ramazanouglu2018spontaneous,ramazanouglu2019spontaneous,ramazanouglu2019generalized,heisenberg2014generalization,garcia2021destabilization} has already been studied for both an Einstein-Maxwell-vector (EMv) model~\cite{oliveira2021spontaneous} and a vector-Gauss-Bonnet (vGB) model~\cite{barton2021spontaneously}\footnote{While vectorized solutions exist in both models, only the EMv has entropically preferable solutions when compared with vacuum GR BHs.}.

    Recent works have, however, cast doubt on the viability of vector fields around astrophysical objects \cite{demirbouga2022instability,silva2022ghost} due to the presence of ghost instabilities~\cite{demirbouga2022instability,unluturk2023loss,coates2023pervasiveness}. In fact, models with vector fields seem to be plagued with ghost instabilities (see \textit{e.g.} \cite{Himmetoglu09} and references therein), with massive vector fields being especially sensitive. Self-interactions have also been shown to originate ghosts in otherwise seemingly ghost-free objects like Proca stars \cite{coates2022intrinsic,clough2022ghost}. 

    One major difference from scalar fields is in the number of additional degrees of freedom. Scalar fields contribute a single new degree of freedom irrespective of whether they are part of scalarization or not -- provided that the scalar field equation is second order in derivatives. However, this is not the case for vectorization. In general, vector-tensor theories break the gauge freedom found in massless vector fields. As a result, the vector field of vectorization models carries three degrees of freedom instead of the two found in electromagnetism.

    This is not immediately a problem, since minimally coupled massive vectors (\textit{aka} Proca), also break the gauge symmetry and still provide a well-behaved classical field theory. However, the extra degree of freedom appears to be problematic with vectorization.
	
	It seems then reasonable to ask if the same also occurs for vectorized BHs. In this work, we present analytical evidence that, in a generic family of extended-vector-tensor theories where the vector field is non-minimally coupled to the model's invariant (source term), a spherically symmetric, vectorized BH always suffers from ghost instabilities. For this, we follow the approach of \cite{clough2022ghost,coates2022intrinsic}, which consists of re-writing the field equations in a wave-like fashion, and we study the behaviour of the effective metric arising from the computation.
	
	Throughout the paper, $4\pi G=1=4\pi\epsilon_0$. The signature of the spacetime is $(-,+,+,+)$. In this work, one is solely interested in spherical symmetry and the metric matter functions are only radially dependent. For notation simplicity, after being first introduced, the functions' radial dependence is omitted, \textit{e.g.} $X(r)\equiv X$, and  $X' \equiv dX/dr$, and we consider the notation $\hat{X}\equiv dX /d \textbf{B}^2$ for the derivative with respect to the vector field.

    The paper is organized as follows. In Sec.~\ref{S2} we review the basic concept of ghost instabilities, while the framework for vectorization is presented in Sec.~\ref{S3}. We then show the occurrence of ghost instabilities in extended-vector-tensor (eVT) theory with a generic source term in Sec.~\ref{S4} and we derive our main conclusions in Sec.~\ref{Sec.Conc}. Appendix \ref{A} and \ref{B} are devoted to the application of the previous result to the EMv and a vGB model, respectively.
%
\section{Basics of instabilities}\label{S2}
%
    In this section, we briefly review the general aspects of ghosts and gradient instabilities in a simplified scenario. For this, let us follow \cite{doneva2022scalarization} and consider the linearized scalar field equation in the $1+1$ dimensional spacetime
    \begin{equation}
        g^{tt}\, \partial _t  ^2 \delta \phi +g^{xx}\, \partial_x ^2 \delta \phi = \mu ^2 \phi\ , 
    \end{equation}
    where, for simplicity, assume a diagonal metric with constant components. The absence of instabilities requires
    \begin{equation}
        g^{tt} < 0\ , \qquad \qquad g^{xx} >0\ , \qquad \qquad \mu ^2 >0\ .
    \end{equation}
   If the field is described by a plain wave mode $\delta \phi (t,x) \propto e^{i(\omega t-k x)}$, the resulting the dispersion relation comes as
    \begin{equation}
        \omega (k) =\sqrt{\frac{\mu ^2 + g^{xx} k^2}{-g^{tt}}}\ .
    \end{equation}
    Three types of instability are then present. For $\mu^2 <0$, the mode behaves as a tachyon, and $\omega (k)$ becomes imaginary for sufficiently small $|k|$, leading to exponential growth. The fastest growing mode behaves as $\sim e^{\sqrt{\mu ^2 /g^{tt}}\,t}$, implying an upper limit to the growth state.

    When $g^{tt}>0$ a ghost instability settles in. There is also an exponential growth of the field, however, this time the rate of growth diverges with increasing wave number as $\sim e^{\sqrt{g^{xx}/g^{tt}}t}$. 

    If $g^{xx}<0$, the same asymptotic behaviour as the ghost exists, creating an instability known as gradient instability. Observe that both ghost and gradient instabilities are qualitatively different from the limited growing tachyonic instability.
%
\section{Vectorization}\label{S3}
%
     As already mentioned in the introduction Sec.~\ref{S1}, while spontaneous vectorized solutions have been studied, when compared to the scalarization phenomena, a huge gap in the literature still exists. In particular, concerning the latter, three types of solutions have been observed to exist \cite{astefanesei2019einstein}: dilatonic, connected scalarization (\textit{aka} linear or spontaneous scalarization) and disconnected scalarization (\textit{aka} non-linear scalarization); while for the former only connected vectorized solutions have been studied\footnote{The dilatonic solution seems to be incompatible with vector fields due to the vector nature of the field, however a term of the kind $\sqrt{|B_\mu B^\nu|}$ could solve the problem. The latter is although not the topic of this work.}.
     
     In this work, we are interested in a class of eVT theories which can be generically described by the action
    \begin{equation}\label{E4}
        \mathcal{S}_{v\mathcal{I}}=\frac{1}{16\pi G} \int d^4 x \sqrt{-g}\Big[ R-G^{\mu\nu} G_{\mu \nu}+f(\textbf{B}^2) \mathcal{I}\Big]\ ,
    \end{equation}
    where $G_{\mu\nu}=\nabla_\mu B_\nu -\nabla_\nu B_\mu$ is the field strength, $f(\textbf{B}^2)$ is a coupling function that couples non-minimally the real vector field to the $\mathcal{I}$ source term which is an invariant of the theory (two examples of $\mathcal{I}$ will be given in Sec.~\ref{S4}). Let us also define the derivative of the coupling function with respect to the vector field $\hat{f}\equiv \frac{d f(\textbf{B}^2)}{dB ^2}$, with $\textbf{B}^2 = B_\mu B^\mu$.
    
    Variation of the action \eqref{E4} with respect to the vector and metric fields gives the corresponding field equations 
    \begin{align}
        \label{eq:proca}
        \nabla _\mu G^{\mu \alpha}&=-\frac{1}{2}\hat{f}\, \mathcal{I}\,B^\alpha\ ,\\
        R_{\mu\nu}-\frac{1}{2}g_{\mu\nu}R &= 2T_{\mu\nu} \label{EFEquations}\ ,
    \end{align}
    with the stress-energy tensor $T_{\mu \nu}$,
    \begin{align}\label{stress}
	T_{\mu\nu} =& f(|B|^2)\bigg(F_\mu^{\,\,\alpha}F_{\nu\alpha} - \frac{1}{4}g_{\mu\nu}F^{\alpha\beta}F_{\alpha\beta}\bigg) + \frac{1}{2}\bigg(G_\mu^{\,\,\alpha}G^*_{\nu\alpha}+G^{*\alpha}_\mu G_{\nu\alpha}-\frac{1}{2}g_{\mu\nu}G^{\mu\nu}G^*_{\mu\nu}\bigg) \nonumber \\
	 &+ \frac{1}{4}\frac{df}{d|B|^2}F^{\alpha\beta}F_{\alpha\beta}\big(B_\mu B^*_\nu+ B^*_\mu B_\nu) \ .
    \end{align}
 
     Observe that, $B_t (r) = 0$ solves the field equations, and thus the vacuum BH is a solution. This requires that 
            \begin{equation}
               \hat{f}(0) \equiv \frac{df(\textbf{B}^2)}{d\textbf{B}^2}\bigg|_{B_t=0} = 0\ ,
            \end{equation}
        which is easily implemented if one requires a $Z_2-$invariance: $B_t \to -B_t$. The vectorized solutions are, however, in general not unique. Vectorization can be separated into two subclasses. To create a parallelism with the scalarized case, let us keep the same notation and consider the vectorization as class II, which is further separated into two subclasses:
    \begin{itemize}
        \item \textbf{Subclass IIA or linearly/normal vectorized type:}
        In this subclass of v-$\mathcal{I}$ models, the vectorized BHs bifurcate from the vacuum BHs and reduce to the latter for $B_t= 0 $. This bifurcation moreover, may be associated with a tachyonic instability against vector (linear) perturbation of the vacuum BH. Let us consider a small-$B_t$ expansion of the coupling function
        \begin{equation}
            f(B_t) = f(0)+\frac{\hat{f}(0)}{2}+\cdots\ .
        \end{equation}
        The linearized Proca equation \eqref{eq:proca} for small $B_t$ reads:
        \begin{equation}
            \nabla _\mu G^{\mu \nu}=-\frac{\hat{f}}{2}\, \mathcal{I}\, B^\nu\ ,
        \end{equation}
        with an effective mass
        \begin{equation}
            \mu_{eff}^2=-\frac{\hat{f}(0)}{2}\,\mathcal{I}\ .
        \end{equation}
        The instability arises if $\mu _{eff}^2<0$.

        \item \textbf{Subclass IIB or non-linearly vectorized type:}
          In this class of vector-$\mathcal{I}$ models, the vectorized BHs \textit{do not} bifurcate from the vacuum BH and \textit{do not} reduce to the later for $B_t = 0$. This is the case if there is no tachyonic instability but there is a non-linear instability. A sufficient (but not necessary) condition is that
            \begin{equation}
                \hat{f}(0) =0\ ,
            \end{equation}
        A non-linear instability implies
            \begin{equation}
             \mathcal{I}\, \frac{d^2 f(\textbf{B}^2)}{d(\textbf{B}^2)^2} \equiv \mathcal{I}\,\hat{\hat{f}} \leqslant 0\ ,
            \end{equation}
        with the difference in the sign associated with the $\textbf{B}^2$ that comes from the first derivative (see \ref{A} and \ref{B} for an example). A mixed vectorization with both mechanisms: tachyonic and nonlinear vectorization is also possible \footnote{Please see \cite{doneva2022beyond} for an example of nonlinear and mixed scalarization in Gauss-Bonnet.}.
    \end{itemize}
%
\section{Vector-$\mathcal{I}$ ghost instability}\label{S4}
%
     To show the presence of the ghost instability, it is important to remember that, in spherical symmetry, $\mathcal{I} \equiv \mathcal{I}(r)$ and that $\textbf{B}^2 (r)$ is negative everywhere outside the event horizon\footnote{The vector field $B_\mu$ of the vectorized BH can be shown to only contain the time component in the static limit: $B_\mu \equiv B_t dt$. This means that, due to our metric signature and assuming $B_t \geqslant 0$ for all the spacetime, $\textbf{B}^2=B_t\, g^{tt}B_t<0$.}. Let us also introduce the scalar quantity
    \begin{equation}
        z = -\frac{1}{2}\hat{f}\,\mathcal{I}\equiv z(r,\textbf{B}^2),
    \end{equation}
    which is a function of both the field and the spacetime coordinates.
    
    In order to identify the condition on $f(\textbf{B}^2)$ (or on $z$) for which ghost instabilities arise, one must write the Proca eq. \eqref{eq:proca} as a wave equation for the vector field $B^\mu $ with an effective "mass" matrix $\mathcal{M}_{\alpha \beta}$. In this regard, one needs to expand the field equation in terms of the vector field $B^\mu$. The resulting Proca equation \eqref{eq:proca} is
    \begin{equation}
     \label{eq:procaex}
     \begin{split}
         0 = & \nabla ^\mu G_{\mu \nu} -z B_\nu\ \\
         = &  \nabla ^\mu \big( \nabla_\mu B_\nu -\nabla _\nu B_\mu \big) -zB_\nu = \nabla ^\mu \nabla_\mu B_\nu-\nabla ^\mu \nabla_\nu B_\mu - z B_\nu \,.
     \end{split}    
    \end{equation}
    By using the definition of the Riemann and Ricci tensors
    \begin{equation}
        R^d _{\ cab}B^c = \nabla_a \nabla_b B^d - \nabla _b \nabla_a B^d\ , \qquad \qquad R_{\mu \nu} = R^c _{\ \mu c \nu}\ ,
    \end{equation}
    the second term in the last equality of \eqref{eq:procaex} can be replaced. This leads to
    \begin{equation}
        \label{eq:procR}
        0= \nabla ^\mu \nabla _\mu B_\nu -\nabla _\nu \nabla _\mu B^\mu -R_{\mu \nu} B^\mu-z B_ \nu\ .
    \end{equation}
    Where the first term is the wave operator acting on the vector field, while the second term $\nabla _\nu \nabla _\mu B^\mu $, needs to be rewritten to render the whole equation manifestly hyperbolic. For this, consider the modified Lorentz condition:
    \begin{equation}
        \nabla_\mu \nabla _\nu G^{\mu \nu} = 0 = \nabla ^\mu (z B_\mu)\,,
    \end{equation}
    which comes from the antisymmetry of the $G^{\mu \nu}$ tensor. This can be rewritten as
    \begin{equation}
        \nabla_\mu \big( zB^\mu \big)  = 0 \Rightarrow \nabla _\mu B^\mu = -\frac{1}{z}B^\mu \nabla _\mu z\ ,
    \end{equation}
    which we further insert in \eqref{eq:procR}. Note that, first-order derivatives of $\nabla _\mu B^\nu$ do not contribute to the dynamics when the vector field is expanded in $B^\mu = B_0 ^\mu + \epsilon \delta B^\mu$ around a constant $B_\mu ^0$, and hence, only the no-derivative and second-order derivative terms matter. Equation \eqref{eq:procR} becomes
    \begin{equation}
        0 = \nabla ^\mu \nabla _\mu B_\nu +\nabla _\nu \left( \frac{1}{z}B^\mu\nabla_\mu z\right) -R_{\mu \nu} B^\mu -z B_\nu \ .
    \end{equation}
    Expanding the derivatives in the second term, we get
    \begin{equation} 
         0 = \nabla ^\mu \nabla _\mu B_\nu +\nabla _\nu B^\mu \frac{\nabla _\mu z}{z}- B^\mu \frac{\nabla _\nu z \nabla _\mu z}{z^2}+B^\mu \frac{\nabla _\nu \nabla _\mu z}{z}-R_{\mu \nu} B^\mu -zB_\nu\ . \label{E1.27}
    \end{equation}
    As shown in \cite{clough2022ghost,coates2022intrinsic}, if the radial dependence of $z$ is solely through $B_\mu (r)$, then \eqref{E1.27} can be rephrased as:
    \begin{equation}
    \label{eq:wavez1}
        \nabla ^\mu \nabla _\mu B_\nu +\Big( \nabla _\mu \ln |z| \Big) \nabla _\nu B^\mu = \mathcal{M}_{\mu \nu} B^\mu\ ,
    \end{equation}
    with the effective mass matrix $\mathcal{M}$:
    \begin{equation}
        \mathcal{M}_{\mu \nu} = -\nabla _\mu \nabla _\nu \ln |z| + R_{\mu \nu}+z g_{\mu \nu}  \ .
    \end{equation}
    However, recall that in the v-$\mathcal{I}$ model the situation is more complicated since $z$ depends both implicitly -- from $B_t(r)$ -- and explicitly -- from $\mathcal{I}$ -- on $r$. Due to the shape of $z$, we can divide it into $z(\textbf{B}^2 , r)\equiv z_B (\textbf{B}^2) \cdot z_r (r) $. With this ansatz, the covariant derivative of $z$ is given by
     \begin{align}
         \nabla _\mu z = & z_r \nabla _\mu\, z _B+z_B\nabla _\mu\,  z_r\nonumber\\
                        = & 2\, z_r\, \hat{z}_B\, B^\nu \nabla _\mu B_\nu+z_B\, z_r '\, \delta _r ^\mu \ ,\label{E1.30}
     \end{align}
    where $\frac{d X(\textbf{B}^2)}{dB_\mu} =2\hat{X} B^\nu$. The second-order derivative is
    \begin{equation}\label{E1.33}
        \nabla _\nu \nabla _\mu z = 2\, z_r\, \hat{z}_B\, B^\alpha \nabla_\nu \nabla _\mu B_\alpha + z_r ''\, z_B\, \delta _{\mu}^{r} \, \delta_{\nu} ^{r}+\mathcal{X}_0\ ,
    \end{equation}
    with all the first-order contributions included into the $\mathcal{X}_0$ term. Observe that the only term containing second-order derivatives of the coupling function $f$ (\textit{i.e.} $\hat{\hat{z}}_B$) is included in the $\mathcal{X}_0$. The other term we need to compute is the product of two derivatives of $z$. This is given by
   \begin{align}
    \label{eq:nablanabla}
        \nabla_\nu z \nabla_\mu z = & \Big(z_B\, z_r'\,\delta_\nu^r+2\, z_r\,\hat{z}_B\, B^\rho\nabla_\nu B_\rho\Big)\left(z_B\, z_r'\,\delta_\mu^r+2\,z_r\,\hat{z}_B\, B^\alpha\nabla_\mu B_\alpha\right) \nonumber \\
        = & z_B^2\, (z_r')^2\, \delta_\mu^r\,\delta_\nu^r+ \mathcal{X}_1\ ,
   \end{align}
    where again, $\mathcal{X}_1$ contains all the terms with first-order derivatives. Introducing the results of \eqref{E1.33} and \eqref{eq:nablanabla} into \eqref{E1.27}, and keeping only the second-order derivative and no-derivative terms, we obtain:
    \begin{equation}
        0=\nabla ^\mu \nabla _\mu B_\nu +\left[\left( \frac{z_B\, z_r''}{z}-\frac{z_B^2\, (z_r')^2}{z^2} \right)\delta_\mu^r\, \delta_\nu^r-R_{\mu\nu}-z\, g_{\mu\nu}\right] B^\mu+2\,\frac{z_r\, \hat{z}_B}{z}B^\alpha\nabla_\nu\nabla_\mu B_\alpha B^\mu\ .        
    \end{equation}
    The above equation can be rewritten in a more handful form by reorganizing the various terms as 
    \begin{align}
        \nabla ^\mu \nabla _\mu B_\nu
        +2\, \frac{\hat{z}_B}{z_B}B^\alpha\nabla_\nu\nabla_\mu B_\alpha B^\mu & = B^\mu \Bigg[\left( \frac{(z_r')^2}{z_r^2}-\frac{z_r''}{z_r}\right)\delta_\mu^r\,\delta_\nu^r+R_{\mu \nu} + z\, g_{\mu \nu}\Bigg]\nonumber\\
        g^{\mu \alpha} \nabla _\mu \nabla _\alpha B_\nu + 2\, B^\mu B^\alpha \Big(\frac{\hat{z}_B}{z_B}\Big) \nabla _\mu \nabla _\alpha B_\nu  + 2\, B^\mu B^\alpha \Big(\frac{\hat{z}_B}{z_B}\Big) \nabla _\mu G_{\nu\alpha} & =
        B^\mu \Bigg[\left( \frac{(z_r')^2}{z_r^2}-\frac{z_r''}{z_r}\right)\delta_\mu^r\, \delta_\nu^r+R_{\mu \nu} + z\, g_{\mu \nu}\Bigg]\ , 
    \end{align}
    where we have used the definition of the field strength tensor in the second term of the \textit{lhs}. This can be further rewritten as a no-derivative term by means of the Proca equation \eqref{eq:proca}. The resulting equation is
    \begin{equation}
     z_r\Big[z_B\, g^{\mu \alpha}+2\,\hat{z}_B B^\mu B^\alpha \Big]\nabla _\mu \nabla _\alpha B_\nu = B^\mu \mathcal{M}_{\nu \mu}\ ,
    \end{equation}
    where
    \begin{align}
     \mathcal{M}_{\alpha \beta} &= z\, g_{\alpha\beta}\left(z-\frac{\hat{z}_B\, z_r}{2}\, \textbf{B}^2 \right) +z\left[R_{\alpha\beta}+\left(\frac{(z_r')^2}{z_r^2}-\frac{z_r''}{z_r}\right)\delta_\alpha^r\delta_\beta^r\right]\ ,\nonumber\\
     \tilde{g}_{\mu \nu} &=  z_r\Big[z_B\, g_{\mu \alpha}+2\,\hat{z}_B B_\mu B_\alpha \Big]\ .
     \end{align}
    The ghost instability appears if the effective metric satisfies $\tilde{g} ^{tt} >0$. The condition for the existence of the ghost instability can be obtained by contracting the $\tilde{g}_{\mu \nu}$ metric with the time-like normal vector $n^\mu$. Decomposing the vector field into the scalar potential $\psi$ and a purely spatial vector $X^\mu$ as $B^\mu = X^\mu + n^\mu \psi$ results in 
    \begin{align}
        \tilde{g}_{nn} &= \tilde{g}_{\mu \nu}n^\mu n^ \nu\nonumber\\
         & = z_r\left[2\,\psi ^2 \, \hat{z}_B -z_B\right]\ .
    \end{align}
    Imposing the ghost condition $\tilde{g}_{nn} \geqslant 0$
    \begin{equation}
        z_r\left[2\,\psi ^2\, \hat{z}_B - z_B\right] \geqslant 0\  \label{E27}
    \end{equation}
    Observe that, we can recover the results obtained in \cite{clough2022ghost,coates2022intrinsic} with $z_B =V(r)$ and $z_r \equiv -\mathcal{I} = 1$, with $V(r)$ the self-interacting potential of the vector field. 

    Note that the ghost instability condition \eqref{E27} can be re-expressed in terms of the $z$ function as
    \begin{equation}
        2\psi^2 \hat{z}-z\geqslant 0\ ,
    \end{equation}
    which can be further expressed in terms of the coupling function and source term $\mathcal{I}$ as
    \begin{equation}
        -\psi ^2\mathcal{I} \hat{\hat{f}}+\frac{\mathcal{I}}{2} \hat{f} \geqslant 0\ . 
    \end{equation}
    As shown in Sec.~\ref{S2}, a tachyonic instability arises when $\frac{\mathcal{I}}{2}\hat{f} >0$, which means that, in the absence of higher order terms, a tachyonic instability of a v-$\mathcal{I}$ model is always followed by a ghost instability.
    
    On the other hand, in the absence of a tachyonic instability, $\hat{f} =0$, non-linear instabilities occur when $\frac{\mathcal{I}}{2}\hat{\hat{f}}< 0$, making them also prone to ghost instabilities. It seems then that, no vectorization is able to endow vectorized solutions that are free of ghosts in a v-$\mathcal{I}$ model. We show two exemplary cases of models that can generate vectorized solutions and for which ghost instabilities seem to exist in the appendix (EMv \ref{A} and GBv \ref{B}).
%
\section{Conclusion}\label{Sec.Conc}
%
    In this work, we have provided evidence that spherically symmetric BHs with vector hair coming from a vectorization process in an eVT theory are always prone to ghost instabilities independently of the functional form of the coupling function between the real vector field and the theory's invariant. 

    We performed analytical calculations suggesting that vectorized BH solutions from extended-vector-tensor theories with a non-minimal coupling between the field and a model's invariant $\mathcal{I}$ are always prone to ghost instabilities. The computation is based on the approach presented in previous studies (\textit{e.g}~\cite{clough2022ghost,coates2022intrinsic}), where the ghost is identified by looking at the effective metric which arises when re-writing the Proca equation in a wave-like form. The method doesn't require any assumption on the specific value of the vector field, which further indicates that the ghost could appear for all possible vectorized configurations. 

    Observe that, in this work, we have only dealt with ghost instabilities associated with the time-time component of the effective metric; one could then assume that a change in the metric signature and/or vector field ansatz could avoid such instabilities. However, the same process can also be performed for the spatial components, for which one expects that a gradient instability will emerge, leaving the model again unstable. This appears to indicate a physical origin of the instabilities.

    In addition, our analysis considered only spherically symmetric solutions. With the addition of the adimensional spin $J$, the process of detecting a ghost instability seems to be simpler due to the change of sign in $z$ for $J>0.5$ in some regions of the spacetime. For all the other spins, a result similar to the one present here is expected. 

    One may argue that, just like a tachyonic instability, non-linearities of the model may be able to tame the exponential growth of vector hair and end up with a dynamically viable solution. Nevertheless, while a tachyonic instability has an upper bound to the growth, a ghost/gradient instability does not, making it harder to quench. In order to provide a definitive statement about the overall stability and viability of the solutions, full numerical time evolution study should be performed. This is, however, beyond the scope of this paper and it will be left for future work.

    Finally, we would like to comment on the possible generalization of the current results. The vectorization mechanism can be seen as a special case of a wider class of phenomena called tensorization~\cite{ramazanouglu2019spontaneous,doneva2022scalarization}. However, all such theories seem to be plagued with ghost instabilities and hence one could assume that the current result may be extended to general tensor fields. It is worth to point out that models with scalar fields, due to the absence of additional degrees of freedom are less prone to instabilities and perhaps the most relevant on the astrophysical point of view.

%
\section*{Acknowledgments}
%
 We would like to thank Daniela Doneva, Nuno M. Santos and Jo\~ao M. S. Oliveira for their valuable discussions and comments. A. M. Pombo is supported by the Czech Grant Agency (GA\^CR) under grant number 21-16583M.

\appendix 
\addcontentsline{toc}{section}{APPENDICES}

 %
\section{Einstein-Maxwell-vector}\label{A}
%
    Let us now apply the main result of the paper to two eVT models which are known to generate spontaneous vectorized solutions: EMv and vGB (appendix~\ref{B}). Consider first the Einstein-Maxwell-vector case where the source term is a ``matter'' source: $\mathcal{I} \equiv F_{\mu \nu} F^{\mu \nu}$, with $A_\mu$ the $4$-vector potential and $F_{\mu \nu} =\partial _\mu  A_{\nu} - \partial _\nu A_{\mu}$ the Maxwell tensor. The resulting $z$ components are,
    \begin{equation}
        z_B =\frac{df(\textbf{B}^2)}{d\textbf{B}^2}\ ,\qquad \qquad {\rm and}\ \qquad \qquad  z_r =-\frac{\mathcal{I}}{2} = -\frac{F_{\mu \nu} F^{\mu \nu} }{2}=\frac{Q^2}{2 r^4}\ .
    \end{equation}
    The onset of instability occurs when a vector field perturbs a vacuum Reissner-Nordstrom BH. The metric line element is
    \begin{equation}\label{EA3}
        ds^2 = -\Big(1-\frac{r_H}{r}\Big)dt^2+\frac{dr^2}{(1-\frac{r_H}{r})}+r^2\big(d\theta ^2+\sin ^2 d\varphi ^2\big)\ ,
    \end{equation}
    with $r\geqslant r_H= M^2+\sqrt{M^2-Q^2}$ the horizon radious of the BH and $Q$ the electric charge. Observe that $z_r\geqslant 0$ for any $r\geqslant r_H$, however, the sign of $z_B$ will depend on its functional form. In the literature, an exponential coupling was considered \cite{oliveira2021spontaneous}. Let us consider the simplest, but generic, polynomial case (all the other functions reduce to the polynomial form for small vector field values)
    \begin{equation}
    \label{eq:coupfuncton}
        f=1+ \mathcal{C}_k \textbf{B}^{2k}\,,
    \end{equation}
    with $k=1,\, 2,\, ...$ an integer. 
    
    For the lowest order(s) ($k=1,\, 2$), a tachyonic instability is settled when $\mathcal{C}_1 < 0$, while a non-linear instability is settled when $\mathcal{C}_2  > 0$ -- the difference in sign comes from the negative sign associated with $\textbf{B}^2$. 
    
    The statement on the sign of the $\mathcal{C}_k$ coefficient can be extended to higher powers, such that an instability leading to a growth of the vector field occurs whenever $\mathcal{C}_k <0\, (>0)$ if $k$ is odd (even).

    The ghost instability appears when 
    \begin{align}
    z_r\Big[ 2\ \psi ^2\, \hat{z}_B-z_B\Big]& \geqslant 0\ ,\nonumber\\
    \frac{Q^2}{2 r^4}\Big[2\,\psi ^2\,\mathcal{C}_k \,k(k-1)\,\textbf{B}^{2(k-2)} -\mathcal{C}_k \,k \,\textbf{B}^{2(k-1)}   \Big]&\geqslant 0\ ,\nonumber\\
    \mathcal{C}_k \,k\,\textbf{B}^{2(k-2)}\Big[2\,\psi ^2\,(k-1) - \textbf{B}^{2} \Big]&\geqslant 0\ . 
     \end{align}   

   Since $\textbf{B}^{2}$ is always negative, the above equation can be rewritten as 
        \begin{align} \label{eq:conditionEMV}
         \mathcal{C}_k \,k\,\textbf{B}^{2(k-2)}\Big[2\,\psi ^2\,(k-1) + |\textbf{B}^{2}| \Big]&\geqslant 0\ ,\nonumber \\
         \mathcal{C}_k \,k\,\textbf{B}^{2(k-2)} &\geqslant 0\ ,
        \end{align}
    where the second inequality comes from the fact that the terms inside square brackets are always positive for non-trivial solutions. When $k$ is even, $ \textbf{B}^{2(k-2)} > 0 $,  and a ghost instability is settled for $\mathcal{C}_k \geqslant 0$. On the other hand, when $k$ is odd $ \textbf{B}^{2(k-2)} < 0 $ and $\mathcal{C}_k \leqslant 0$. As a consequence, there is no vectorized solution to EMv black holes free of ghosts.

    A set of exemplary solutions of non-linear vectorized BHs in EMv models have been computed. It was observed that solutions do exist and are entropically preferable when compared with vacuum solutions, however, the study and analysis of such solutions is not the point of the current work. We leave such an exercise for a future paper.
 
%
\section{Einstein-Gauss-Bonnet-vector}\label{B}
%
    In the vector-Gauss-Bonnet model case, the source term is a geometric source: $\mathcal{I}=R_{GB} ^2 \equiv R^2 -4R_{\mu \nu}R^{\mu \nu}+R_{\mu \nu \rho \delta}R^{\mu \nu \rho \delta}$ the Gauss-Bonnet scalar. $\mathcal{I}\equiv \mathcal{G}$,
    \begin{equation}
        z_B =\frac{df(\textbf{B}^2)}{d\textbf{B}^2}\ ,\qquad \qquad {\rm and}\ \qquad \qquad  z_r  = -\frac{\mathcal{G}}{2}=-\frac{24 M^2}{r^6}\ ,
    \end{equation}
    where we have assumed a Schwarzschild background with $r\geqslant r_H= 2M$ the horizon radius of the BH in the line element \eqref{EA3}. 
    
    Observe that $z_r\leqslant 0$ for any $r\geqslant r_H$, however, the sign of $z_B$ will depend on its functional form. In the literature, three terms have been considered \cite{barton2021spontaneously}. Let us use the same polynomial expansion as before (the constant term is absent in agreement with the GBv theory) 
    \begin{equation}
        f=\mathcal{C}_k \textbf{B}^{2k}\,,
    \end{equation}
    The condition for the coefficients  $\mathcal{C}_k$ is now reversed, \textit{i.e.} an instability leading to a growth of the vector field occurs whenever $\mathcal{C}_k >0\, (<0)$ if $k$ is odd (even). This is due to the opposite sign of the $z_r$, which is reflected in the behaviour of the field through the Proca equation \eqref{eq:proca}.

    The condition for ghost instabilities becomes
    \begin{align}
        -\frac{24 M^2}{r^6}\Big[2\,\psi ^2\,\mathcal{C}_k \,k(k-1)\,\textbf{B}^{2(k-2)} -\mathcal{C}_k \,k \,\textbf{B}^{2(k-1)}   \Big]&\geqslant 0\ ,\nonumber\\
        \Big[2\,\psi ^2\,\mathcal{C}_k \,k(k-1)\,\textbf{B}^{2(k-2)} -\mathcal{C}_k \,k \,\textbf{B}^{2(k-1)}   \Big]&\leqslant 0\ .      
    \end{align}

    Applying the same reasoning as before,
    
    \begin{align} \label{eq:conditionGBV}
        \mathcal{C}_k \,k\,\textbf{B}^{2(k-2)}\Big[2\,\psi ^2\,(k-1) + |\textbf{B}^{2}| \Big]&\leqslant 0\ ,\nonumber \\
        \mathcal{C}_k \,k\,\textbf{B}^{2(k-2)} &\leqslant 0\ ,
    \end{align}

    So, a ghost instability is settled for $\mathcal{C}_k \geqslant 0\, (\leqslant 0)$ for $k$ odd (even). Thus, also in this case, all the fully-vectorized solutions of a Schwarzschild BH with a Gauss-Bonnet invariant are affected by ghost instabilities.

\bibliographystyle{ieeetr} 
\bibliography{master}

\begin{thebibliography}{10}

\bibitem{abbott2016observation}
B.~P. Abbott, R.~Abbott, T.~Abbott, M.~Abernathy, F.~Acernese, K.~Ackley, C.~Adams, T.~Adams, P.~Addesso, R.~Adhikari, {\em et~al.}, ``Observation of gravitational waves from a binary black hole merger,'' {\em Physical review letters}, vol.~116, no.~6, p.~061102, 2016.

\bibitem{abbott2021gwtc}
R.~Abbott, T.~Abbott, S.~Abraham, F.~Acernese, K.~Ackley, A.~Adams, C.~Adams, R.~Adhikari, V.~Adya, C.~Affeldt, {\em et~al.}, ``Gwtc-2: compact binary coalescences observed by ligo and virgo during the first half of the third observing run,'' {\em Physical Review X}, vol.~11, no.~2, p.~021053, 2021.

\bibitem{collaboration2019first}
E.~H.~T. Collaboration, K.~Akiyama, A.~Alberdi, W.~Alef, K.~Asada, R.~AZULY, {\em et~al.}, ``First m87 event horizon telescope results. i. the shadow of the supermassive black hole,'' {\em Astrophys. J. Lett}, vol.~875, no.~1, p.~L1, 2019.

\bibitem{akiyama2022first}
K.~Akiyama, A.~Alberdi, W.~Alef, J.~C. Algaba, R.~Anantua, K.~Asada, R.~Azulay, U.~Bach, A.-K. Baczko, D.~Ball, {\em et~al.}, ``First sagittarius a* event horizon telescope results. i. the shadow of the supermassive black hole in the center of the milky way,'' {\em The Astrophysical Journal Letters}, vol.~930, no.~2, p.~L12, 2022.

\bibitem{Tuna22}
S.~{Tuna}, K.~I. {{\"U}nl{\"u}t{\"u}rk}, and F.~M. {Ramazano{\v{g}}lu}, ``{Constraining scalar-tensor theories using neutron star mass and radius measurements},'' {\em \prd}, vol.~105, p.~124070, June 2022.

\bibitem{khodadi2020black}
M.~Khodadi, A.~Allahyari, S.~Vagnozzi, and D.~F. Mota, ``Black holes with scalar hair in light of the event horizon telescope,'' {\em Journal of Cosmology and Astroparticle Physics}, vol.~2020, no.~09, p.~026, 2020.

\bibitem{berti2015testing}
E.~Berti, E.~Barausse, V.~Cardoso, L.~Gualtieri, P.~Pani, U.~Sperhake, L.~C. Stein, N.~Wex, K.~Yagi, T.~Baker, {\em et~al.}, ``Testing general relativity with present and future astrophysical observations,'' {\em Classical and Quantum Gravity}, vol.~32, no.~24, p.~243001, 2015.

\bibitem{Kobayashi:2018xvr}
T.~Kobayashi and T.~Hiramatsu, ``{Relativistic stars in degenerate higher-order scalar-tensor theories after GW170817},'' {\em Phys. Rev.}, vol.~D97, no.~10, p.~104012, 2018.

\bibitem{Burrage19}
C.~{Burrage}, J.~{Dombrowski}, and D.~{Saadeh}, ``{The shape dependence of Vainshtein screening in the cosmic matter bispectrum},'' {\em jcap}, vol.~2019, p.~023, Oct. 2019.

\bibitem{Pizzuti22}
L.~{Pizzuti}, I.~D. {Saltas}, K.~{Umetsu}, and B.~{Sartoris}, ``{Probing vainsthein-screening gravity with galaxy clusters using internal kinematics and strong and weak lensing},'' {\em mnras}, vol.~512, pp.~4280--4290, May 2022.

\bibitem{Ballardini23}
M.~{Ballardini}, A.~G. {Ferrari}, and F.~{Finelli}, ``{Phantom scalar-tensor models and cosmological tensions},'' {\em jcap}, vol.~2023, p.~029, Apr. 2023.

\bibitem{crisostomi2016extended}
M.~Crisostomi, K.~Koyama, and G.~Tasinato, ``Extended scalar-tensor theories of gravity,'' {\em Journal of Cosmology and Astroparticle Physics}, vol.~2016, no.~04, p.~044, 2016.

\bibitem{doneva2022scalarization}
D.~D. Doneva, F.~M. Ramazano{\u{g}}lu, H.~O. Silva, T.~P. Sotiriou, and S.~S. Yazadjiev, ``Scalarization,'' {\em arXiv preprint arXiv:2211.01766}, 2022.

\bibitem{antoniou2018evasion}
G.~Antoniou, A.~Bakopoulos, and P.~Kanti, ``Evasion of no-hair theorems and novel black-hole solutions in gauss-bonnet theories,'' {\em Physical review letters}, vol.~120, no.~13, p.~131102, 2018.

\bibitem{doneva2018new}
D.~D. Doneva and S.~S. Yazadjiev, ``New gauss-bonnet black holes with curvature-induced scalarization in extended scalar-tensor theories,'' {\em Physical review letters}, vol.~120, no.~13, p.~131103, 2018.

\bibitem{silva2018spontaneous}
H.~O. Silva, J.~Sakstein, L.~Gualtieri, T.~P. Sotiriou, and E.~Berti, ``Spontaneous scalarization of black holes and compact stars from a gauss-bonnet coupling,'' {\em Physical review letters}, vol.~120, no.~13, p.~131104, 2018.

\bibitem{herdeiro2018spontaneous}
C.~A. Herdeiro, E.~Radu, N.~Sanchis-Gual, and J.~A. Font, ``Spontaneous scalarization of charged black holes,'' {\em Physical review letters}, vol.~121, no.~10, p.~101102, 2018.

\bibitem{doneva2022dynamical}
D.~D. Doneva, A.~Va{\~n}{\'o}-Vi{\~n}uales, and S.~S. Yazadjiev, ``Dynamical descalarization with a jump during a black hole merger,'' {\em Physical Review D}, vol.~106, no.~6, p.~L061502, 2022.

\bibitem{gibbons1975vacuum}
G.~W. Gibbons, ``Vacuum polarization and the spontaneous loss of charge by black holes,'' {\em Communications in Mathematical Physics}, vol.~44, pp.~245--264, 1975.

\bibitem{fernandes2019spontaneous}
P.~G. Fernandes, C.~A. Herdeiro, A.~M. Pombo, E.~Radu, and N.~Sanchis-Gual, ``Spontaneous scalarisation of charged black holes: coupling dependence and dynamical features,'' {\em Classical and Quantum Gravity}, vol.~36, no.~13, p.~134002, 2019.

\bibitem{astefanesei2019einstein}
D.~Astefanesei, C.~Herdeiro, A.~Pombo, and E.~Radu, ``Einstein-maxwell-scalar black holes: classes of solutions, dyons and extremality,'' {\em Journal of High Energy Physics}, vol.~2019, no.~10, pp.~1--27, 2019.

\bibitem{fernandes2019charged}
P.~G. Fernandes, C.~A. Herdeiro, A.~M. Pombo, E.~Radu, and N.~Sanchis-Gual, ``Charged black holes with axionic-type couplings: Classes of solutions and dynamical scalarization,'' {\em Physical Review D}, vol.~100, no.~8, p.~084045, 2019.

\bibitem{herdeiro2021aspects}
C.~A. Herdeiro, A.~M. Pombo, and E.~Radu, ``Aspects of gauss-bonnet scalarisation of charged black holes,'' {\em Universe}, vol.~7, no.~12, p.~483, 2021.

\bibitem{brihaye2019spontaneous}
Y.~Brihaye and B.~Hartmann, ``Spontaneous scalarization of charged black holes at the approach to extremality,'' {\em Physics Letters B}, vol.~792, pp.~244--250, 2019.

\bibitem{jiang2020spontaneous}
S.~Jiang, ``Spontaneous scalarization of charged gauss-bonnet black holes: Analytic treatment,'' {\em arXiv preprint arXiv:2011.03998}, 2020.

\bibitem{myung2019instability}
Y.~S. Myung and D.-C. Zou, ``Instability of reissner--nordstr{\"o}m black hole in einstein-maxwell-scalar theory,'' {\em The European Physical Journal C}, vol.~79, pp.~1--11, 2019.

\bibitem{myung2019quasinormal}
Y.~S. Myung and D.-C. Zou, ``Quasinormal modes of scalarized black holes in the einstein--maxwell--scalar theory,'' {\em Physics Letters B}, vol.~790, pp.~400--407, 2019.

\bibitem{BlazquezSalcedo:2020nhs}
J.~L. Bl\'azquez-Salcedo, C.~A.~R. Herdeiro, J.~Kunz, A.~M. Pombo, and E.~Radu, ``{Einstein-Maxwell-scalar black holes: the hot, the cold and the bald},'' {\em Phys. Lett. B}, vol.~806, p.~135493, 2020.

\bibitem{LuisBlazquezSalcedo:2020rqp}
J.~Luis Bl\'azquez-Salcedo, C.~A.~R. Herdeiro, S.~Kahlen, J.~Kunz, A.~M. Pombo, and E.~Radu, ``{Quasinormal modes of hot, cold and bald Einstein\textendash{}Maxwell-scalar black holes},'' {\em Eur. Phys. J. C}, vol.~81, no.~2, p.~155, 2021.

\bibitem{doneva2022beyond}
D.~D. Doneva and S.~S. Yazadjiev, ``Beyond the spontaneous scalarization: New fully nonlinear mechanism for the formation of scalarized black holes and its dynamical development,'' {\em Physical Review D}, vol.~105, no.~4, p.~L041502, 2022.

\bibitem{pombo2023effects}
A.~M. Pombo and D.~D. Doneva, ``Effects of mass and self-interaction on nonlinear scalarization of scalar-gauss-bonnet black holes,'' {\em arXiv preprint arXiv:2310.08638}, 2023.

\bibitem{annulli2022spin}
L.~Annulli, C.~A. Herdeiro, and E.~Radu, ``Spin-induced scalarization and magnetic fields,'' {\em Physics Letters B}, vol.~832, p.~137227, 2022.

\bibitem{hod2020onset}
S.~Hod, ``Onset of spontaneous scalarization in spinning gauss-bonnet black holes,'' {\em Physical Review D}, vol.~102, no.~8, p.~084060, 2020.

\bibitem{dima2020spin}
A.~Dima, E.~Barausse, N.~Franchini, and T.~P. Sotiriou, ``Spin-induced black hole spontaneous scalarization,'' {\em Physical Review Letters}, vol.~125, no.~23, p.~231101, 2020.

\bibitem{herdeiro2021spin}
C.~A. Herdeiro, E.~Radu, H.~O. Silva, T.~P. Sotiriou, and N.~Yunes, ``Spin-induced scalarized black holes,'' {\em Physical review letters}, vol.~126, no.~1, p.~011103, 2021.

\bibitem{berti2021spin}
E.~Berti, L.~G. Collodel, B.~Kleihaus, and J.~Kunz, ``Spin-induced black hole scalarization in einstein-scalar-gauss-bonnet theory,'' {\em Physical Review Letters}, vol.~126, no.~1, p.~011104, 2021.

\bibitem{collodel2020spinning}
L.~G. Collodel, B.~Kleihaus, J.~Kunz, and E.~Berti, ``Spinning and excited black holes in einstein-scalar-gauss--bonnet theory,'' {\em Classical and Quantum Gravity}, vol.~37, no.~7, p.~075018, 2020.

\bibitem{ramazanouglu2017spontaneous}
F.~M. Ramazano{\u{g}}lu, ``Spontaneous growth of vector fields in gravity,'' {\em Physical Review D}, vol.~96, no.~6, p.~064009, 2017.

\bibitem{ramazanouglu2018spontaneous}
F.~M. Ramazano{\u{g}}lu, ``Spontaneous growth of gauge fields in gravity through the higgs mechanism,'' {\em Physical Review D}, vol.~98, no.~4, p.~044013, 2018.

\bibitem{ramazanouglu2019spontaneous}
F.~M. Ramazano{\u{g}}lu, ``Spontaneous tensorization from curvature coupling and beyond,'' {\em Physical Review D}, vol.~99, no.~8, p.~084015, 2019.

\bibitem{ramazanouglu2019generalized}
F.~M. Ramazano{\u{g}}lu and K.~{\.I}. {\"U}nl{\"u}t{\"u}rk, ``Generalized disformal coupling leads to spontaneous tensorization,'' {\em Physical Review D}, vol.~100, no.~8, p.~084026, 2019.

\bibitem{heisenberg2014generalization}
L.~Heisenberg, ``Generalization of the proca action,'' {\em Journal of Cosmology and Astroparticle Physics}, vol.~2014, no.~05, p.~015, 2014.

\bibitem{garcia2021destabilization}
S.~Garcia-Saenz, A.~Held, and J.~Zhang, ``Destabilization of black holes and stars by generalized proca fields,'' {\em Physical Review Letters}, vol.~127, no.~13, p.~131104, 2021.

\bibitem{oliveira2021spontaneous}
J.~M. Oliveira and A.~M. Pombo, ``Spontaneous vectorization of electrically charged black holes,'' {\em Physical Review D}, vol.~103, no.~4, p.~044004, 2021.

\bibitem{barton2021spontaneously}
S.~Barton, B.~Hartmann, B.~Kleihaus, and J.~Kunz, ``Spontaneously vectorized einstein-gauss-bonnet black holes,'' {\em Physics Letters B}, vol.~817, p.~136336, 2021.

\bibitem{demirbouga2022instability}
E.~S. Demirbo{\u{g}}a, A.~Coates, and F.~M. Ramazano{\u{g}}lu, ``Instability of vectorized stars,'' {\em Physical Review D}, vol.~105, no.~2, p.~024057, 2022.

\bibitem{silva2022ghost}
H.~O. Silva, A.~Coates, F.~M. Ramazano{\u{g}}lu, and T.~P. Sotiriou, ``Ghost of vector fields in compact stars,'' {\em Physical Review D}, vol.~105, no.~2, p.~024046, 2022.

\bibitem{unluturk2023loss}
K.~{\.I}. {\"U}nl{\"u}t{\"u}rk, A.~Coates, and F.~M. Ramazano{\u{g}}lu, ``Loss of hyperbolicity and tachyons in generalized proca theories,'' {\em arXiv preprint arXiv:2306.03554}, 2023.

\bibitem{coates2023pervasiveness}
A.~Coates and F.~M. Ramazano{\u{g}}lu, ``Pervasiveness of the breakdown of self-interacting vector field theories,'' {\em Physical Review D}, vol.~107, no.~10, p.~104036, 2023.

\bibitem{Himmetoglu09}
B.~{Himmetoglu}, C.~R. {Contaldi}, and M.~{Peloso}, ``{Ghost instabilities of cosmological models with vector fields nonminimally coupled to the curvature},'' {\em \prd}, vol.~80, p.~123530, Dec. 2009.

\bibitem{coates2022intrinsic}
A.~Coates and F.~M. Ramazano{\u{g}}lu, ``Intrinsic pathology of self-interacting vector fields,'' {\em Physical Review Letters}, vol.~129, no.~15, p.~151103, 2022.

\bibitem{clough2022ghost}
K.~Clough, T.~Helfer, H.~Witek, and E.~Berti, ``Ghost instabilities in self-interacting vector fields: The problem with proca fields,'' {\em Physical review letters}, vol.~129, no.~15, p.~151102, 2022.

\end{thebibliography}

\end{document}